# Quantized Hall conductance in graphene by nonperturbative magnetic-field-containing relativistic tight-binding approximation method

Md. Abdur Rashid[1,2], Masahiko Higuchi[3] and Katsuhiko Higuchi[1*]

[1]Graduate School of Advanced Science and Engineering, Hiroshima University, Higashi-Hiroshima 739-8527, Japan
[2]Department of Physics, University of Rajshahi, Rajshahi 6205, Bangladesh
[3]Departments of Physics, Faculty of Science, Shinshu University, Matsumoto 390-8621, Japan
*khiguchi@hiroshima-u.ac.jp

**ABSTRACT**

In this study, we conducted a numerical investigation on the Hall conductance ($\sigma_{Hall}$) of graphene based on the magnetic energy band structure calculated using a nonperturbative magnetic-field-containing relativistic tight-binding approximation (MFRTB) method. The nonperturbative MFRTB can revisit two types of plateaus for the dependence of $\sigma_{Hall}$ on Fermi energy. One set is characterized as wide plateaus (WPs). These WPs have filling factors (FFs) of 2, 6, 10, 14, etc. and are known as the half-integer quantum Hall effect. The width of WPs decreases with increasing FF, which exceeds the decrease expected from the linear dispersion relation of graphene. The other set is characterized by narrow plateaus (NPs), which have FFs of 0, 4, 8, 12, etc. The NPs correspond to the energy gaps caused by the spin-Zeeman effect and spin-orbit interaction. Furthermore, it was discovered that the degeneracy of the magnetic energy bands calculated using the nonperturbative MFRTB method leads to a quantized $\sigma_{Hall}$.

## I. INTRODUCTION

Graphene is a monolayer of carbon atoms and demonstrates certain peculiar properties for its energy band diagram, that is, a linear energy dispersion relationship near the Dirac point. Graphene, with its distinctive properties, has garnered considerable attention as a promising material in the electronics and spintronics fields [1-8]. Graphene immersed in a uniform magnetic field also exhibits distinctive properties such as strong orbital diamagnetism [9-22], reduced effective g-factor [23-25], unconventional oscillation of magnetization [26-28], and half-integer quantum Hall effect (QHE) [29-42]. Graphene is becoming a more practical material for QHE-based applications at room temperature because of the large energy gap generated by the magnetic field compared to a standard two-dimensional electron gas (2DEG) system [41].

Half-integer QHE in graphene has attracted scientific attention since its theoretical predictions [29,30] and experimental discoveries [2,31,39-42]. Quantized Hall conductance ($\sigma_{Hall}$) with filling factors (FFs) of 2, 6, 10, 14, etc. was experimentally observed in a low magnetic field of approximately 14 (T) [2] or below 10 (T) [31], which is known as the half-integer QHE. The half-integer QHE was theoretically described based on the tight-binding (TB) approximation method and/or the effective mass Hamiltonian method [30,32-38]. In addition to the half-integer QHE, quantized $\sigma_{Hall}$ with FFs of 0 and 4 was also observed experimentally in the high magnetic field region of approximately 45 (T) [39]. The quantized $\sigma_{Hall}$ with FFs of 0, 4, 8, 12, etc. has also been described theoretically and attributed to the energy splitting caused by the spin-Zeeman effect [37].

The theoretically obtained magnetic-field dependence of the energy levels of electrons in a magnetic field is often referred to as the Hofstadter butterfly diagram [43]. The gap in the Hofstadter butterfly diagram provides information on QHEs [44-47]. In conventional theoretical methods [30,32-38], this diagram is calculated using hopping integrals multiplied by the Peierls phase as magnetic hopping integrals (hopping integrals in the presence of a magnetic field). However, as the Peierls phase approximation corresponds to a lowest-order perturbation theory [48], it becomes evident that the diagram is incorrect in the high magnetic field region and lacks accuracy even in the low magnetic field region [48]. Furthermore, the Hofstadter butterfly diagram in the low magnetic field region is likely to be affected by spin-orbit interactions [48]. Therefore, it would be more appropriate to consider both the nonperturbative effects of the magnetic field and the spin-orbit interaction in investigating the QHE.

We recently developed the magnetic-field-containing relativistic tight-binding approximation (MFRTB) [49] and the nonperturbative MFRTB methods [48] to describe the properties of materials immersed in a uniform magnetic field. These methods enable the calculation of the realistic energy band structure of materials immersed in a magnetic field (magnetic energy band structure) by considering the effects of the magnetic field, periodic potential, and relativity. Thus far, the MFRTB method can revisit the Haas-van Alphen oscillations [50,51] and magnetic breakdown [51] and predict the additional oscillation peaks of the magnetization [52-54]. The nonperturbative MFRTB method reveals that nonperturbative effects appear in high and low magnetic field regions [48] and successfully predicts the second-order phase transition of silicon from a band insulator to a metal [55]. All these discussions are based on the magnetic energy band structure calculated using the MFRTB or nonperturbative MFRTB method.

In this study, we investigate the QHE in graphene based on the magnetic energy band structure calculated using the nonperturbative MFRTB method. In particular, we investigate the Fermi energy dependence of $\sigma_{Hall}$ using the nonperturbative MFRTB method. The nonperturbative MFRTB method can revisit the quantized $\sigma_{Hall}$ with FFs of 2, 6, 10, etc. and with 0, 4, 8, etc., where the former has wide plateaus (WPs) and the latter has narrow plateaus (NPs). As shown later, the width of WPs decreases with an increase in the FF owing to the negative curvature of the energy band for a zero magnetic field. The width of NPs is shown to be determined by the spin-Zeeman effect and the spin-orbit interaction.

The remainder of this paper is organized as follows. In Section II, we explain the calculation method for the magnetic energy band structure and quantized $\sigma_{Hall}$. In Section IIIA, we present the magnetic energy band structure of graphene and discuss the degeneracy. In Sections III B and C, the Hofstadter butterfly diagram calculated using the nonperturbative MFRTB method is presented. We also discuss the Fermi energy dependence of the width of WPs and NPs in Section III D. In Section III E, we demonstrate that the degeneracy of the magnetic energy band leads to a quantized $\sigma_{Hall}$. Finally, the conclusions are presented in Sec. IV.

**II. CALCULATION METHOD**

The nonperturbative MFRTB method [48] was applied to graphene immersed in a magnetic field. The outline of the nonperturbative MFRTB method is given in Appendix. Specific formulas for the Hamiltonian matrix elements used in the application of the nonperturbative MFRTB method to graphene, and other details, are found in Ref. 22. The magnetic field was assumed to be perpendicular to the plane of graphene, and its magnitude is given by

$$B = \frac{8\pi\hbar}{\sqrt{3}ea^2}\frac{p}{q},\quad (1)$$

where $a$ (24.6 $nm$) denotes the lattice constant of graphene and $p$ and $q$ are relatively prime integers [22]. Note that the magnetic field, the magnitude of which is proportional to the rational number $p/q$, is sometimes called the rational magnetic field [43, 56, 57]. Following previous studies [22, 27, 36, 38, 43, 48, 49, 53-62], the calculations in this study were performed under the assumption of a rational magnetic field [63]. In the case of graphene, the rational magnetic field is given by Eq. (1). In the present calculations, the magnetic hopping integrals between the outer shells (2s- and 2p-orbitals) of the nearest-neighbor carbon atoms were considered. Their values were calculated using a table of nonperturbative magnetic hopping integrals (Table I of Ref. [48]) and the relativistic version of the Slater–Koster table [49]. We adopted a set of relativistic tight-binding parameters for graphene that is given in the previous paper [48]. The magnetic energy band structure of graphene was calculated for the wave vectors lying in the MBZ [49]. The MBZ of graphene immersed in the magnetic field of Eq. (1) is illustrated in Fig. 1.

Hall conductance was calculated based on the obtained magnetic energy band structure. Specifically, the number of states below the Fermi energy ($\varepsilon_F$) was calculated from the magnetic energy band structure, and $\sigma_{Hall}$ was calculated by substituting it into the Streda formula [47]:

$$\sigma_{Hall} = e\frac{\partial n(B)}{\partial B},\quad (2)$$

where $n(B)$ denotes the number of states below $\varepsilon_F$.

### III. RESULTS AND DISCUSSION

A. Magnetic energy band structure

To investigate the quantized $\sigma_{Hall}$ based on the magnetic energy band structure, we first discuss the features of the magnetic energy band of graphene that is calculated using the nonperturbative MFRTB method [48]. Figures 2(a) and 2(b) show the magnetic Bloch bands of graphene in the presence of magnetic fields with $B = 200.5$ (T) and $B' = 200.9$ (T), respectively, and the values of $p/q$ are 1/787 and 2/1571, respectively. The aim of considering two close magnetic field values is to investigate the degeneracy of the magnetic energy band, which will be discussed in the next paragraph. First, we mention the features of the magnetic energy band that are common for these two cases. The horizontal axis denotes the wave vector in the MBZ, and the vertical axis denotes the energy in electron volt (eV). From Figs. 2(a) and 2(b), we conclude that the magnetic energy bands of graphene are nearly independent of the values of $k_x$ and $k_y$ in the MBZ. As shown in reference [53], a cluster of nearly flat magnetic energy bands in close energy proximity corresponds to the so-called Landau level that is attributed to Onsager's quantization of the electron orbit. Because the nonperturbative MFRTB method is based on the Dirac equation for an electron moving in both a uniform magnetic field and the periodic potential of the crystal, the spin-orbit interaction and Zeeman effect are inherently considered. Thus, as shown in Figs. 2(a) and 2(b), each flat band splits into two energy bands owing to the spin-orbit interaction and the Zeeman effect.

Next, we discuss the degeneracy of the magnetic energy band. In this study, we introduce the approximation that a nearly flat band is completely flat. In other words, each eigenvalue at Γ point is $N_k$-fold degenerate, where $N_k$ is the number of $k$ points in the MBZ. If the area of the unit cell for the zero magnetic field case is denoted by $A_{unit}$ and if the total number of unit cells contained in the system is denoted by $N$, then the total area of the system is given by $A = A_{unit} N$. Because the area of the magnetic unit cell is $q$-times larger than $A_{unit}$ [22, 48, 49], the total number of magnetic unit cells in the system is given by $N/q$. The total number of magnetic unit cells contained in the system is found to be equal to that of the $k$-points in the MBZ, $N_k$ [49]. Consequently, under this approximation, each eigenvalue at Γ point is $N/q$-fold degenerate.

Subsequently, we introduce another approximation. For this purpose, we considered the magnetic energy band structures for two close magnetic fields, $B$ and $B'$, as shown in Figs. 2(a) and 2(b), respectively. Suppose $B$ and $B'$ are proportional to $p/q$ and $1/q'$, respectively, where $q'$ is a prime integer. The relationship $B \approx B'$ implies that $q \approx pq'$. For this relationship $q \approx pq'$, the period in the real space for the case of $B$ is $p$ times larger than that for the case of $B'$. If the difference of the Hamiltonian for these two cases is treated as the perturbation potential, and if the Hamiltonian for the case of $B'$ corresponds to the nonperturbative Hamiltonian, then the perturbation potential would be small owing to the relationship $B \approx B'$. This small perturbation potential makes the periodicity $p$ times longer than that for the nonperturbative system because of $q \approx pq'$. If we consider the change of the periodicity without considering the small shift in energy due to the small perturbation potential, the magnetic energy bands for the case of $B$ can be obtained by $p$-times folding of the magnetic energy bands for the case of $B'$ within the MBZ of the nonperturbative system. Therefore, we introduce an approximation in which only the change in periodicity is considered without considering the small shift in energy caused by the small perturbation potential. Because of the flatness of the magnetic energy bands, $p$-times folding results in an energy band with $p$-fold degeneracy. This $p$-fold degeneracy was confirmed by actual calculations using the nonperturbative MFRTB method. Figures 2(a) and 2(b) illustrate the case of $B \approx B'$. From these figures, we can confirm that the degeneracy of the magnetic energy band for $p/q$ (=2/1571) is twice that of the magnetic energy band for $1/q'$ (=1/787). For instance, while the eigenvalue around 0 (eV) at $\Gamma$ point is 4-fold degeneracy for $1/q' = 1/787$ (Fig. 2(a)), that is, 8-fold degeneracy for $p/q = 2/1571$ (Fig. 2(b)). Based on the above discussion, we can conclude that the degeneracy of the magnetic energy band is given by $Np/q$.

In addition to the aforementioned degeneracy, it was found from the magnetic energy band calculated using the nonperturbative MFRTB method that the degeneracy of the magnetic energy band is a multiple of four. This degeneracy is due to two reasons. First, each unit cell of graphene in a honeycomb structure consists of two carbon atoms. Therefore, we utilized twice the number of bases to expand the magnetic Bloch function [22, 48, 49]. This leads to two-fold degeneracy. Second, Dirac points exist at two inequivalent points, namely the $K$ and $K'$ points, in the BZ. Consequently, the degeneracy of the magnetic energy band of graphene becomes a multiple of four.

Thus, the degeneracy of the magnetic energy band, $g(B)$, for graphene is given by $g(B) = 4Np/q$. If the area of the system is denoted by $A$, then $N = A/A_{unit}$. Therefore, we have

$$g(B) = 4\,\frac{A}{A_{unit}}\frac{p}{q}. \tag{3}$$

Using Eq. (1), Eq. (3) can be rewritten as follows:

$$g(B) = \frac{2eA\,B}{h}. \tag{4}$$

Equation (4) corresponds to the conventional degeneracy of the Landau level given in the literature [51, 64]. The important point is that $g(B)$ is proportional to the magnetic field. Equation (4) is used later to discuss Hall conductance.

B. Magnetic-field dependence of energy levels

Figure 3 shows the magnetic-field dependence of energy levels of electrons in graphene (Hofstadter butterfly diagram) at $\Gamma$ point that is calculated by the nonperturbative MFRTB method. Figure 4 shows a magnified view of Fig. 3. In Figs. 3 and 4, the horizontal axis denotes the magnetic field in Tesla (T) and the vertical axis denotes the energy in eV. It is well known that the energy spectrum of graphene is proportional to the square root of the magnetic field at low magnetic fields [9,10]. This square root behavior was first predicted by McClure [9,10] based on Onsager's area quantization rule. As shown in Fig. 3, the nonperturbative MFRTB method can revisit the square-root behavior [22]. In addition, owing to the relativistic effects, including the spin-Zeeman effect and spin-orbit interaction, each quantized energy level is split into two energy levels, as shown in Fig. 4. Thus, we observe two types of energy splitting in the magnetic-field-dependent energy spectrum of graphene. The first type is related to Onsager's area quantization rule and is indicated by (i), (ii), and (iii) in Fig. 4. The other is due to relativistic effects and is indicated by (iv), (v), and (vi) in Fig. 4. By comparing energy splitting in (i), (ii), and (iii), it is found that the former decreases with the energy for a constant magnetic field ((i) and (ii)) and increases with the magnetic field ((i) and (iii)). With respect to the latter energy splitting, it is small compared to the former and increases with an increase in the magnetic field ((iv) and (v)). As confirmed later, these two types of energy splitting are the origins of the wide and narrow plateaus in the Fermi energy dependence of the quantized $\sigma_{Hall}$.

C. Quantized Hall conductance by nonperturbative MFRTB method

The quantized $\sigma_{Hall}$ in graphene is calculated using the Streda formula (Eq. (2)) [47] based on the magnetic energy band structure. Figures 5(a)–5(f) show the dependence of $\sigma_{Hall}$ on the position of the Fermi energy for (a) $B = 10.00$ (T), (b) $B = 20.02$ (T), (c) $B = 48.50$ (T), (d) $B = 101.50$ (T), (e) $B = 239.50$ (T), and (f) $B = 600.50$ (T), respectively. Two types of plateaus are observed in these figures. One set of plateaus has a comparatively wide width, with FFs of 2, 6, 10, etc. Another set of plateaus has a comparatively narrow width, with FFs of 0, 4, 8, etc. For all magnetic fields, we confirmed that the starting and ending energies of each plateau in Figs. 5(a)–5(f) correspond exactly to the energy levels shown in Fig. 4. The widths of the plateaus shown in Figs. 5(a)–5(f) are consistent with the energy splitting shown in Fig. 4. The WPs in Figs. 5(a)–5(f) correspond to the energy splitting related to Onsager's area-quantization rule (Fig. 4). In addition, we can confirm that the NPs in Figs. 5(a)–5(f) correspond to energy splitting caused by relativistic effects (Fig. 4). Thus, the quantized $\sigma_{Hall}$ in graphene can be effectively described based on the magnetic energy band structure.

D. Fermi energy dependence of widths of plateaus

As discussed in Section III C, the WP corresponds to the energy splitting related to Onsager's area quantization rule. Figure 6 shows the dependence of the width of WPs on the FF. As shown in Fig. 6, the width of WPs decreases with increasing FF. This is expected to be due to the fact that the energy separation of two successive quantized energy levels caused by Onsager's area quantization rule is not same for all energy levels (for whole energy region) in graphene, unlike in a conventional 2DEG system. To analyze the dependence of the width of WPs on the FF, we calculated the energy separation of two successive quantized energy levels using Onsager's area quantization rule. Based on the area quantization rule, the quantized energy level is given by $E_n = \sqrt{2ev_F^2\hbar|n|B}$, where $n$ denotes the Landau level index, and $v_F$ $(= 1.0 \times 10^6 \, m/s)$ is the Fermi velocity of graphene [9,10]. The gap energy between two successive energy levels is given by

$$\Delta E_n = \sqrt{2ev_F^2\hbar B}\left(\sqrt{|n|+1} - \sqrt{|n|}\right). \tag{5}$$

As mentioned in Section III C, the energy separation given by Eq. (5), corresponds to the width of WPs. The width of WPs calculated using Eq. (5) is shown in Fig. 6 concurrently with those of the nonperturbed MFRTB method. The width of WPs, calculated using Eq. (5), agrees with those obtained using the nonperturbative MFRTB method in the lower FF region, that is, in the low-energy region. This means that Eq. (5) is a good approximation in the low-energy region. Recalling that Eq. (5) is obtained using Onsager's area quantization rule when assuming a linear energy–dispersion relationship at zero magnetic fields, the good agreement suggests that these treatments are valid in the low-energy region, that is, in the lower-FF region. However, a discrepancy appears in the higher-FF region, that is, in the higher-energy region. This is because the assumption of linear energy–dispersion relationship becomes less appropriate as the energy level moves away from the Dirac point. Based on the energy band structure in the absence of a magnetic field, the curvature of the energy dispersion becomes negative as the energy level moves away from the Dirac point. Therefore, the density of the quantized energy levels that satisfy Onsager's area quantization rule increases with the energy range. Therefore, the width of WPs decreases with increasing FF or energy range, which exceeds the decrease expected from the linear energy–dispersion relationship.

Subsequently, we discuss the origin of the NPs in detail. Figure 7 shows the magnetic-field dependence of the width of NPs for FF=4. For reference, the width of the NP expected from the spin-Zeeman effect alone is also indicated in this figure. If the spin-orbit interaction is neglected, the width of NPs coincides with the width expected from the spin-Zeeman effect. Therefore, the difference between the results by the nonperturbative MFRTB method and those by the spin-Zeeman effect, which is denoted as $\Delta$, may be regarded as the effect of the spin-orbit interaction. Figure 8 shows the magnetic-field dependence of the difference $\Delta$. As shown in Figs. 7 and 8, the width of the NP agrees with the width expected from the spin-Zeeman effect in both the low and high magnetic field regions. In the high magnetic field region, the effects of the spin-orbit interaction become negligible compared to the spin-Zeeman effect, which is regarded as the Paschen–Back effect. Owing to the Paschen–Back effect, the width of NPs calculated by the nonperturbative MFRTB method approaches those calculated by the spin-Zeeman effect. The agreement in the low-magnetic-field region indicates that the energy splitting caused by the anomalous Zeeman effect is consistent with that of the spin-Zeeman effect for magnetic Bloch states related to an FF of four. This is possible, for example, when the magnetic quantum number of the total angular momentum is given by $\pm 1/2$, the energy splitting caused by the anomalous Zeeman effect is consistent with that caused by the spin-Zeeman effect in a low magnetic field [48]. Therefore, the agreement in the low magnetic field region implies that the magnetic Bloch states related to FF of four mainly comprise atomic orbitals with the magnetic quantum number of $\pm 1/2$.

On the other hand, the discrepancy is relatively large in the middle magnetic field region (approximately 200 (T)) in Figs. 7 and 8. This implies that the effect of the spin-orbit interaction can be observed in the width of NPs in a magnetic field of approximately 200 (T). Although the width of NPs has not necessarily been estimated accurately in experiments [39], it is expected that the effect of the spin-orbit interaction and Paschen–Back effect will be observed in NPs by further experiments, especially in high magnetic fields greater than 100 (T) [65]. It should be noted that the description of Delta can be realized owing to the nonperturbative treatments of effects of the magnetic field and spin-orbit interaction through the nonperturbative MFRTB method.

E. Description of the quantized Hall conductance based on the magnetic energy band structure

In this section, we explain the reason why $\sigma_{Hall}$ in graphene can be described based on the magnetic energy band structure calculated by the nonperturbative MFRTB method. From Section III A, we confirmed that all the magnetic energy bands approximately have a degeneracy of $g(B)$, that is proportional to the magnetic field (Eq. (4)). If $N(\varepsilon_F)$ is the number of flat magnetic energy bands below the Fermi energy, then the total number of states per unit area below the Fermi energy is given as follows:

$$n(B) = \frac{N(\varepsilon_F)\, g(B)}{A}$$
$$= \frac{2e\, B}{h} N(\varepsilon_F). \tag{6}$$

According to the Streda formula [47], $\sigma_{Hall}$ can be obtained as follows:

$$\sigma_{Hall} = e \frac{\partial}{\partial B} n(B)$$
$$= \frac{2e^2}{h} N(\varepsilon_F). \tag{7}$$

The Hall conductance is proportional to the number of flat magnetic energy bands below the Fermi energy. As $N(\varepsilon_F)$ is an integer, Eq. (7) indicates that $\sigma_{Hall}$ is quantized.

Subsequently, we discuss the relationship between the magnetic energy band structure and $\sigma_{Hall}$ in more detail. The nonperturbative MFRTB provides $16q$ eigenvalues for each $\boldsymbol{k}$-point in the MBZ. Because the total number of magnetic unit cells is given by $N/q$, we obtain $16N$ states using the nonperturbative MFRTB method. In the honeycomb lattice structure of graphene, each magnetic unit cell consists $2q$ number of carbon atoms. Therefore, the magnetic unit cell consists $8q$ number of valence electrons. Therefore, the total number of valence electrons in the system is equal to $8N$ $(= 8q \times N/q)$. For intrinsic graphene, half of the $16N$ states, that is, the $8N$ states, are occupied. The flat band approximation described in Section III A, $8q$ of $16q$ magnetic energy bands are fully occupied. As mentioned in Section III A, each cluster of magnetic energy bands consists of $4p$ magnetic energy bands, which leads to an approximate $p$-fold degeneracy. Considering that the cluster corresponds to the so-called Landau level, we shall assign the following numbering to the clusters: That is, we shall refer to the 0-th cluster that contains $(8q - 4p)$, $(8q - 4p + 1), \cdots, (8q - 1), 8q$-th magnetic energy bands from the bottom. Similarly, the first cluster contains the $(8q + 1), (8q + 2), \cdots, (8q + 4p)$-th magnetic energy bands from the bottom. Herein, the energy level of the nth cluster is denoted by $E_n$.

For the intrinsic graphene, the Fermi energy level lies between the 0-th and 1st cluster, that is, $E_0 < \varepsilon_F < E_1$. Here, the total number of states below the Fermi energy is $8N$. Therefore, the total number of states per unit area is

$$n(B) = \frac{8N}{A}$$
$$= \frac{16}{\sqrt{3}a^2} \quad \text{(for } E_0 < \varepsilon_F < E_1\text{),} \tag{8}$$

where we used the relation $A/N = A_{unit} = \sqrt{3}a^2/2$. Therefore, $n(B)$ is independent of $B$ when $E_0 < \varepsilon_F < E_1$. This leads to

$$\sigma_{Hall} = 0 \quad \text{(for } E_0 < \varepsilon_F < E_1\text{).} \tag{9}$$

Subsequently, we consider the case in which the Fermi energy is located between $E_1$ and $E_2$, that is, $E_1 < \varepsilon_F < E_2$. Here, the total number of states below the Fermi energy is given as $(8 + 4p/q)N \ (= (8q + 4p) \times N/q)$. Therefore, the total number of states per unit area is

$$n(B) = \frac{\left(8 + \frac{4p}{q}\right)N}{A}$$
$$= \frac{16}{\sqrt{3}\, a^2} + \frac{2e}{h} B \ (\text{for } E_1 < \varepsilon_F < E_2), \qquad (10)$$

where we used Eq. (1). Therefore, we have

$$\sigma_{Hall} = \frac{2e^2}{h} \quad (\text{for } E_1 < \varepsilon_F < E_2). \qquad (11)$$

Similarly, we can consider the case in which the Fermi energy lies between the (*n*-1)th and *n*-th clusters. That is, $E_{n-1} < \varepsilon_F < E_n$. Here, the total number of states below the Fermi energy level is $(8 + 4np/q)N \ (= (8q + 4np) \times (N/p))$. Therefore, the total number of states per unit area is

$$n(B) = \frac{\left(8 + \frac{4n\,p}{q}\right)N}{A}$$
$$= \frac{16}{\sqrt{3}\, a^2} + \frac{2e}{h} nB \quad (\text{for } E_{n-1} < \varepsilon_F < E_n). \qquad (12)$$

Therefore, we have

$$\sigma_{Hall} = \frac{e^2}{h} 2n \quad (\text{for } E_{n-1} < \varepsilon_F < E_n). \qquad (13)$$

Thus, $\sigma_{Hall}$ in graphene can be described based on the magnetic energy band structure calculated using the nonperturbative MFRTB method.

**IV. CONCLUSION**

Using the nonperturbative MFRTB method, we investigated the quantized $\sigma_{Hall}$ in graphene. It was confirmed that WPs with FFs of 2, 6, 10, 14, etc. and NPs with FFs of 0, 4, 8, 12, etc. were revisited in the Fermi energy dependence of $\sigma_{Hall}$. The former is attributed to energy splitting, which corresponds to Onsager's area-quantization rule. The latter arises because of the energy splitting caused not only by the spin-Zeeman effect but also by the spin-orbit interaction in a magnetic field.

The width of WPs decreases with increasing Fermi energy for a constant magnetic field, which agrees with the experimental results. The dependence of the width of WPs in the lower-energy region is consistent with the conventional theoretical model [6-9], in which the linear energy dispersion relationship is employed as an approximation for the energy band structure in the absence of a magnetic field. On the other hand, a discrepancy between the conventional theoretical model and the nonperturbative MFRTB method appears in the higher-energy region. Although the conventional theoretical model is valid only close to the Dirac point, the nonperturbative MFRTB method provides a practical magnetic energy band structure that is valid even in the high-energy region. That is, the discrepancy appears owing to the lack of validity in using the linear energy dispersion relationship in the higher-energy region. Thus, the present description of the Fermi energy dependence of the width of WPs is reliable in both the lower and higher energy regions.

It is possible to observe the effect of the spin-orbit interaction and Pachen–Back effect in graphene by investigating the Fermi energy dependence of the width of NPs at magnetic fields greater than 100 (T). In general, the Pachen–Back effect appears if the ratio of the Zeeman splitting to the spin-orbit splitting exceeds one [48]. For graphene, the magnetic field that makes ratio 1 is approximately 144 (T) [48]. This magnetic field appears to be consistent with the results shown in Figs. 7 and 8. Thus, the detection of the Paschen–Back effect in NPs is expected to be realized using the recent progress [66] in generating an extremely high magnetic field greater than 1000 (T) and in measuring physical quantities in the extremely high magnetic field [67-69].

Furthermore, the relationship between the magnetic energy band structure and quantized $\sigma_{Hall}$ was determined. Each time the Fermi energy crosses a cluster of magnetic energy bands, $\sigma_{Hall}$ changes by $2e^2/h$. This statement is similar to the description of dHvA oscillations based on the magnetic energy band structure [52-54].

It would be interesting to consider the effect of the fine energy band structure of a cluster on $\sigma_{Hall}$. This is because the fine energy band structure in a cluster generates additional oscillation peaks in the magnetic oscillation of the dHvA effect [53] under a high magnetic field. The conventional MFRTB and Hofstadter methods are based on perturbation theory [48]; thus, they are inadequate for describing such phenomena. Only the nonperturbative MFRTB method may be applied as a first-principles calculation method for high magnetic field region. The effect of the fine energy band structure will be investigated in future studies.

In the nonperturbative MFRTB method, the electron-electron interactions are not incorporated. The effects of the electron-electron interaction can be incorporated into effective potentials using the density functional theory [70,71] and its extended theories [72-82]. For example, we can use current density functional theory (CDFT) [77-82]. The nonperturbative MFRTB method can be employed to solve the Kohn–Sham (KS) equation of CDFT. It proves intriguing to delve into the effects of electron-electron interaction on the magnetic energy band structure of graphene as well as its implications on the alterations observed in the quantized Hall conductance. This is one of the future issues to be addressed.

**Acknowledgments**

This work was partially supported by Grants-in-Aid for Scientific Research (Nos. 18K03510, 18K03461 and 23K03250) from the Japan Society for the Promotion of Science. This work was supported by JST SPRING (grant number JPMJSP2132).

**Appendix: Outline of nonperturbative MFRTB method**

In this appendix, we mention the outline of the nonperturbative MFRTB method [48]. Let us consider an electron moving in both a uniform magnetic field and periodic potential of the crystal. The Dirac equation for the electron is given by

$$\left[ c\boldsymbol{\alpha} \cdot \{\boldsymbol{p} + e\boldsymbol{A}(\boldsymbol{r})\} + \beta mc^2 + \sum_{\boldsymbol{R}_n} \sum_i V_{a_i}(\boldsymbol{r} - \boldsymbol{R}_n - \boldsymbol{d}_i) \right] \Phi_{\alpha \boldsymbol{k}}(\boldsymbol{r}) = E_\alpha(\boldsymbol{k}) \Phi_{\alpha \boldsymbol{k}}(\boldsymbol{r}), \quad \text{(A1)}$$

where $\boldsymbol{A}(\boldsymbol{r})$ and $V_{a_i}(\boldsymbol{r} - \boldsymbol{R}_n - \boldsymbol{d}_i)$ are the vector potential for an applied uniform magnetic field $\boldsymbol{B} = (0,0,B)$ and scalar potential caused by the nucleus of an atom $a_i$ located at $\boldsymbol{R}_n + \boldsymbol{d}_i$, respectively. The vectors $\boldsymbol{R}_n$ and $\boldsymbol{d}_i$ denote the translational vector of the lattice and vector specifying the position of atom $a_i$, respectively. The letters *c*, *e* and *m* represent the velocity of light, elementary charge and rest mass of an electron, respectively. The quantities $\boldsymbol{\alpha}\,(=(\alpha_x, \alpha_y, \alpha_z))$ and $\beta$ stand for the usual $4 \times 4$ matrices. The eigenfunction $\Phi_{\alpha k}(\boldsymbol{r})$ represents the four components wave function, the subscripts of which denote the band index α and the wave vector $\boldsymbol{k}$ belonging to the Magnetic first Brillouin zone (MBZ).

In the nonperturbative MFRTB method [48], $\Phi_{\alpha k}(\boldsymbol{r})$ is expanded by using the relativistic atomic orbitals immersed in a magnetic field as basis functions;

$$\Phi_{\alpha k}(\boldsymbol{r}) = \sum_{\boldsymbol{R}_n} \sum_i \sum_\xi C_k^\xi(\boldsymbol{R}_n + \boldsymbol{d}_i)\, \psi_\xi^{a_i, \boldsymbol{R}_n + \boldsymbol{d}_i}(\boldsymbol{r}) E_\alpha(\boldsymbol{k}) \Phi_{\alpha k}(\boldsymbol{r}), \qquad (A2)$$

where $\psi_\xi^{a_i, \boldsymbol{R}_n + \boldsymbol{d}_i}(\boldsymbol{r})$ denotes the relativistic atomic orbital for atom $a_i$ located at $\boldsymbol{R}_n + \boldsymbol{d}_i$ that is immersed in a uniform magnetic field $\boldsymbol{B}$. Hereafter we refer to $\psi_\xi^{a_i, \boldsymbol{R}_n + \boldsymbol{d}_i}(\boldsymbol{r})$ as the magnetic atomic orbital. The magnetic atomic orbital obeys the following Dirac equation;

$$[c\boldsymbol{\alpha} \cdot \{\boldsymbol{p} + e\boldsymbol{A}(\boldsymbol{r})\} + \beta mc^2 + V_{a_i}(\boldsymbol{r} - \boldsymbol{R}_n - \boldsymbol{d}_i)] \psi_\xi^{a_i, \boldsymbol{R}_n + \boldsymbol{d}_i}(\boldsymbol{r}) = \varepsilon_\xi^{a_i, \boldsymbol{R}_n + \boldsymbol{d}_i} \psi_\xi^{a_i, \boldsymbol{R}_n + \boldsymbol{d}_i}(\boldsymbol{r}), \quad (A3)$$

where $\varepsilon_\xi^{a_i, \boldsymbol{R}_n + \boldsymbol{d}_i}$ and the subscript ξ denote the atomic spectrum of an atom $a_i$ located at $\boldsymbol{R}_n + \boldsymbol{d}_i$ that is immersed in a uniform magnetic field and the quantum number, respectively. The generalized eigenvalue problem for the expansion coefficient $C_{\alpha k}^\xi(\boldsymbol{R}_n + \boldsymbol{d}_i)$ is obtained from Eq. (A1);

$$\sum_{R_n} \sum_i \sum_\xi H_{R_m j \eta, R_n i \xi} C^\xi_{\alpha k}(R_n + d_i)$$
$$= E_\alpha(k) \sum_{R_n} \sum_i \sum_\xi S_{R_m j \eta, R_n i \xi} C^\xi_{\alpha k}(R_n + d_i), \tag{A4}$$

where $H_{R_m j \eta, R_n i \xi}$ and $S_{R_m j \eta, R_n i \xi}$ represents the matrix elements of the Hamiltonian of Eq. (A1) and inner product between $\psi_\eta^{a_j, R_m + \vec{d}_j}(r)$ and $\psi_\xi^{a_i, R_n + \vec{d}_i}(r)$, respectively. If we neglect integrals involving three different centers, then we have the following expressions for $H_{R_m j \eta, R_n i \xi}$ and $S_{R_m j \eta, R_n i \xi}$;

$$H_{R_m j \eta, R_n i \xi} = (\varepsilon_\xi^{a_i, 0} + \Delta\varepsilon_\xi^{a_i, d_i}) \delta_{j,i} \delta_{\eta,\xi} \delta_{R_m, R_n}$$
$$+ (1 - \delta_{j,i} \delta_{R_m, R_n}) e^{-i\frac{eB}{\hbar}(R_{nx}+d_{ix}-R_{mx}-d_{jx})(R_{my}+d_{jy})}$$
$$\times \left\{ T^{a_j a_i}_{\eta \xi}(R_{nm} + d_i - d_j) + \frac{\varepsilon_\eta^{a_j, 0} + \varepsilon_\xi^{a_i, 0}}{2} S^{a_j a_i}_{\eta \xi}(R_{nm} + d_i - d_j) \right\} \tag{A5}$$

and

$$S_{R_m j \eta, R_n i \xi} = e^{-i\frac{eB}{\hbar}(R_{nx}+d_{ix}-R_{mx}-d_{jx})(R_{my}+d_{jy})} S^{a_j a_i}_{\eta \xi}(R_{nm} + d_i - d_j), \tag{A6}$$

with

$$T^{a_j a_i}_{\eta \xi}(R_{nm} + d_i - d_j)$$
$$= \int \psi_\eta^{a_j, 0}(r) \frac{V_{a_j}(r) + V_{a_i}(r - R_{nm} - d_i + d_j)}{2} \psi_\xi^{a_j, R_{nm}+d_i-d_j}(r) d^3r \tag{A7}$$

and

$$S_{\eta\,\xi}^{a_j a_i}(\boldsymbol{R}_{nm} + \boldsymbol{d}_i - \boldsymbol{d}_j) = \int \psi_\eta^{a_j,\boldsymbol{0}}(\boldsymbol{r})\, \psi_\xi^{a_j,\boldsymbol{R}_{nm}+\boldsymbol{d}_i-\boldsymbol{d}_j}(\boldsymbol{r}) d^3r, \tag{A8}$$

where $\boldsymbol{R}_{nm}$ denotes $\boldsymbol{R}_n - \boldsymbol{R}_m$. In Eq, (A5), $\Delta\varepsilon_\xi^{a_i,\boldsymbol{d}_i}$ is the effect of the crystal field in the presence of the magnetic field. Since $\Delta\varepsilon_\xi^{a_i,\boldsymbol{d}_i}$ is expected to be much smaller than the atomic spectrum, we neglect $\Delta\varepsilon_\xi^{a_i,\boldsymbol{d}_i}$ in the actual calculations [22, 48]. In Eqs. (A7) and (A8), $T_{\eta\,\xi}^{a_j a_i}(\boldsymbol{R}_{nm} + \boldsymbol{d}_i - \boldsymbol{d}_j)$ and $S_{\eta\,\xi}^{a_j a_i}(\boldsymbol{R}_{nm} + \boldsymbol{d}_i - \boldsymbol{d}_j)$ denote the magnetic hopping integral and magnetic overlap integral, respectively [48].

To calculate $H_{\boldsymbol{R}_m j\,\eta,\ \boldsymbol{R}_n i\,\xi}$ and $S_{\boldsymbol{R}_m j\,\eta,\ \boldsymbol{R}_n i\,\xi}$, we need values of $T_{\eta\,\xi}^{a_j a_i}(\boldsymbol{R}_{nm} + \boldsymbol{d}_i - \boldsymbol{d}_j)$, $S_{\eta\,\xi}^{a_j a_i}(\boldsymbol{R}_{nm} + \boldsymbol{d}_i - \boldsymbol{d}_j)$ and $\varepsilon_\xi^{a_i,\ \boldsymbol{R}_n+\boldsymbol{d}_i}$. These can be obtained using the magnetic atomic orbital. In the nonperturbative MFRTB method, the magnetic atomic orbital is approximately estimated by means of the variational method [48]. Specifically, the magnetic atomic orbital is expanded in terms of the relativistic atomic orbital in the absence of a magnetic field [48]. By substituting the expansion into Eq. (A3), we obtain the approximate form of the magnetic atomic orbital and the atomic spectrum. The specific forms of the resultant magnetic atomic orbital and atomic spectrum are given in Refs. [22, 48]. Since the resultant magnetic atomic orbital is expressed in the linear combination of the relativistic atomic orbital, $T_{\eta\,\xi}^{a_j a_i}(\boldsymbol{R}_{nm} + \boldsymbol{d}_i - \boldsymbol{d}_j)$ and $S_{\eta\,\xi}^{a_j a_i}(\boldsymbol{R}_{nm} + \boldsymbol{d}_i - \boldsymbol{d}_j)$ can be expressed by the linear combination of relativistic hopping integrals and overlap integrals in the absence of a magnetic field. The resultant approximate forms for $T_{\eta\,\xi}^{a_j a_i}(\boldsymbol{R}_{nm} + \boldsymbol{d}_i - \boldsymbol{d}_j)$ and $S_{\eta\,\xi}^{a_j a_i}(\boldsymbol{R}_{nm} + \boldsymbol{d}_i - \boldsymbol{d}_j)$ are given in Table II of Ref [48]. Note that relativistic hopping integrals and overlap integrals in the absence of a magnetic field can be expressed in terms of several relativistic TB parameters [49], which is summarized in the relativistic version of the Slater-Koster table (Table I of Ref. [49]). Thus, we can calculate the matrix elements of $H_{R_m j \eta, R_n i \xi}$ and $S_{R_m j \eta, R_n i \xi}$ from several relativistic TB parameters. Relativistic TB parameters are determined by requiring that energy-band structure for the zero magnetic field case a calculated by the relativistic TB approximation method [49] coincides with that of the reference data as well as possible. As the reference data for graphene, the energy-band structure calculated by wien2k code [83] with taking the spin-orbit interaction into account [22]. This is an outline of the nonperturbative MFRTB method.

**References**

[1] K. S. Novoselov, A. K. Geim, S. V. Morozov, D. Jiang, Y. Zhang, S. V. Dubonos, I. V. Grigorieva, and A. A. Firsov, Sci. **306**, 666 (2004).

[2] K. S. Novoselov, A. K. Geim, S. V. Morozov, D. Jiang, M. I. Katsnelson, I. V. Grigorieva, S. V. Dubonos, and A. A. Firsov, Nat. **438**,197 (2005).

[3] A. K. Geim and K. S. Novoselov, Nature Mat. **6**, 183 (2007).

[4] P. Avouris, T. F. Heinz, and T. Low, *2D Materials: Properties and Devices* (Cambridge University Press, Cambridge, 2017)

[5] K. Higuchi, M. Kudo, M. Mori, and T. Mishima, IEDM Tech. Dig. 891 (1994).

[6] M. Balkanski, *Devices Based on Low-Dimensional Semiconductor Structures* (Kluwer Academic Publishers, Dordrecht, 1996)

[7] K. Higuchi, H. Matsumoto, T. Mishima, and T. Nakamura, IEEE Trans. Elec. Dev. **46**, 1312 (1999).

[8] G. Hellings and K. De Myer, *High Mobility and Quantum Well Transistors* (Springer, New York, 2013)

[9] J. W. McClure, Phys. Rev. **104**, 666 (1956).

[10] J. W. McClure, Phys. Rev. **119**, 606 (1960).

[11] S. A. Safran and F. J. DiSalvo, Phys. Rev. B **20**, 4889 (1979).

[12] S. A. Safran, Phys. Rev. B **30**, 421 (1984).

[13] R. Saito and H. Kamimura, Phys. Rev. B **33**, 7218 (1986).

[14] H. Fukuyama, J. Phys. Soc. Jpn. **76**, 043711 (2007).

[15] M. Koshino and T. Ando, Phys. Rev. B **75**, 235333 (2007).

[16] M. Koshino, Y. Arimura, and T. Ando, Phys. Rev. Lett. **102**, 177203 (2009).

[17] M. Sepioni, R. R. Nair, S. Rablen, J. Narayanan, F. Tuna, R. Winpenny, A. K. Geim, and I. V. Grigorieva, Phys. Rev. Lett. **105**, 207205 (2010).

[18] G. Gómez-Santos and T. Stauber, Phys. Rev. Lett. **106**, 045504 (2011).

[19] A. Raoux, F. Píechon, J. -N. Fuchs, and G. Montambaux, Phys. Rev. B **91**, 085120 (2015).

[20] Y. Gao, S. A. Yang, and Q. Niu, Phys. Rev. B **91**, 214405 (2015).

[21] M. Ogata and H. Fukuyama, J. Phys. Soc. Jpn. **84**, 124708 (2015).

[22] M. Higuchi, D. B. Hamal, A. Shrestha, and K. Higuchi, J. Phys. Soc. Jpn. **88**, 094707 (2019).

[23] R. G. Mani, J. Hankinson, C. Berger, and W. A. de Heer, Nat. Commun. **3,** 996 (2012).

[24] T. J. Lyon, J. Sichau, A. Dorn, A. Centeno, A. Pesquera, A. Zurutuza, and R. H. Blick, Phys. Rev. Lett. **119**, 066802 (2017).

[25] A. Shrestha, K. Higuchi, S. Yoshida, and M. Higuchi, J. Appl. Phys. **130**, 124303 (2021)

[26] S. G. Sharapov, V. P. Gusynin, and H. Beck, Phys. Rev. B **69**, 075104 (2004).

[27] K. Kishigi and Y. Hasegawa, Phys. Rev. B **90**, 085427 (2014).

[28] F. Escuderoa, J. S. Ardenghi, L. Sourrouille, and P. Jasen, J. Mag. Mag. Mater. **429**, 294 (2017).

[29] F. D. M. Haldane, Phys. Rev. Lett. **61**, 2015 (1988).

[30] Y. Zheng and T. Ando, Phys. Rev. B **65**, 245420 (2002).

[31] Y. Zhang, Y.-W. Tan, H. L. Störmer, and P. Kim, Nat. (London) **438**, 201 (2005).

[32] V. P. Gusynin and S. G. Sharapov, Phys. Rev. Lett. **95**, 146801 (2005).

[33] V. P. Gusynin and S. G. Sharapov, Phys. Rev. B **73**, 245411 (2006).

[34] N. M. R. Peres, F. Guinea, and A. H. Castro Neto, Phys. Rev. B **73**, 125411 (2006).

[35] D. N. Sheng, L. Sheng, and Z. Y. Weng, Phys. Rev. B **73**, 233406 (2006).

[36] Y. Hasegawa and M. Kohmoto, Phys. Rev. B **74**, 155415 (2006).

[37] V. P. Gusynin, V. A. Miransky, S. G. Sharapov, and I. A. Shovkovy, Phys. Rev. B **74**, 195429 (2006).

[38] Y. Hatsugai, T. Fukui, and H. Aoki, Phys. Rev. B **74**, 205414 (2006).

[39] Y. Zhang, Z. Jiang, J. P. Small, M. S. Purewal, Y.-W. Tan, M. Fazlollahi, J. D. Chudow, J. A. Jaszczak, H. L. Stormer, and P. Kim, Phys. Rev. Lett. **96,** 136806 (2006).

[40] Z. Jiang, Y. Zhang, Y.-W. Tan, J. A. Jaszczak, H. L. Stormer, and P. Kim, Int. J. Mod. Phys. B **21**, 1123 (2007).

[41] K. S. Novoselov, Z. Jiang, Y. Zhang, S. V. Morozov, H. L. Stormer, U. Zeitler, J. C. Maan, G. S. Boebinger, P. Kim, and A. K. Geim, Sci. **315,** 1379 (2007).

[42] X. Wu, Y. Hu, M. Ruan, N. K. Madiomanana, J. Hankinson, M. Sprinkle, C. Berger, and W. A. de Heer, Appl. Phys. Lett. **95**, 223108 (2009).

[43] D. Hofstadter, Phys. Rev. B **14**, 2239 (1976).

[44] D. J. Thouless, M. Kohmoto, M. P. Nightingale, and M. denNijs, Phys. Rev. Lett. **49**, 405 (1982).

[45] M. Kohmoto, Ann. Phys. **160**, 343 (1985).

[46] A. H. MacDonald, Phys. Rev. B **28**, 6713 (1983).

[47] P. Streda, J. Phys. C **15**, L1299 (1982).

[48] K. Higuchi, D. B. Hamal, and M. Higuchi, Phys. Rev. B **97**, 195135 (2018).

[49] K. Higuchi, D. B. Hamal, and M. Higuchi, Phys. Rev. B **91**, 075122 (2015).

[50] W. J. de Haas and P. M. van Alphen, Proc. Netherland R. Acad. Sci. **33**, 680 (1930).

[51] D. Shoenberg, *Magnetic Oscillation in Metals* (Cambridge University Press, Cambridge, 1984).

[52] D. B. Hamal, M. Higuchi, and K. Higuchi, Phys. Rev. B **91**, 245101 (2015).

[53] M. Higuchi, D. Bahadur Hamal, and K. Higuchi, Phys. Rev. B **95**, 195153 (2017).

[54] K. Higuchi, D. B. Hamal, and M. Higuchi, Phys. Rev. B **96**, 235125 (2017).

[55] K. Higuchi, D. B. Hamal, and M. Higuchi, New J. Phys. **24** 103028 (2022).

[56] E. Brown, Phys. Rev. **133**, A1038 (1964).

[57] J. Zak, Phys. Rev. **134**, A1602 (1964).

[58] Y. Hasegawa, Y. Hatsugai and M. Kohmoto, Phys. Rev. B **41**, 9174 (1990).

[59] M. Koshino, H. Aoki and K. Kuroki, Phys. Rev. Lett. **86**, 1062 (2001).

[60] V. M. Gvozdikov and M. Taut, Phys. Rev. B **75**, 155436 (2007).

[61] G.Möller and N.R. Cooper, Phys. Rev. Lett. **115**, 126401 (2015).

[62] R. Huber, M.-N. Steffen, M. Drienovsky, A. Sandner, K. Watanabe, T. Taniguchi, D. Pfannkuche, D. Weiss and J. Eroms, Nat. Commun. **13**, 2856 (2022).

[63] The rational magnetic field given by Eq. (1) is the magnitude of the magnetic field such that the order of the Abelian group composed of magnetic translation operators is maximal and such that the solution of the nonperturbed MFRTB method satisfies the magnetic Bloch theorem [22].

[64] J. Kübler, *Theory of Itinerant Electron Magnetism* (Oxford University Press, New York, 2000), Chap. 1.

[65] The generation of more than 1000 Tesla is realized in laboratories by means of the electromagnetic flux compression method [66]. In addition, using this technique, experimental measurements of physical quantities are successfully done up to about 600 Tesla [67-69].

[66] D. Nakamura, A. Ikeda, H. Sawabe, Y. H. Matsuda, and S. Takeyama, Rev. Sci. Instrum. **89**, 095106 (2018).

[67] Y. H. Matsuda, D. Nakamura, A. Ikeda, S. Takeyama, Y. Suga, H. Nakahara, and Y. Muraoka, Nat. Commun. **11**, 3591 (2020).

[68] D. Nakamura, H. Saito, H. Hibino, K. Asano, and S. Takeyama, Phys. Rev. B **101**, 115420 (2020).

[69] A. Ikeda, Y. H. Matsuda, K. Sato, Y. Ishii, H. Sawabe, D. Nakamura, S. Takeyama and J. Nasu, Nat. Commun. **14**, 1744 (2023).

[70] P. Hohenberg and W. Kohn, Phys. Rev. **136**, B864 (1964).

[71] W. Kohn and L. J. Sham, Phys. Rev. **140**, A1133 (1965).

[72] M. Higuchi and K. Higuchi, Phys. Rev. B **69**, 035113 (2004).

[73] K. Bencheikh and G. Vignale, Phys. Rev. B **77** 155315 (2008).

[74] M. Higuchi and K. Higuchi, J. Phys. Condens. Matter **21**, 064206 (2009).

[75] K. Higuchi and M. Higuchi, Phys. Rev. B. **82**, 155135 (2010).

[76] M. Higuchi and K. Higuchi, Comput. Theor. Chem. **1003**, 91 (2013).

[77] G. Vignale and M. Rasolt, Phys. Rev. Lett. **59**, 2360 (1987).

[78] G. Vignale and M. Rasolt, Phys. Rev. B **37**, 10685 (1988).

[79] K. Higuchi and M. Higuchi, J. Phys. Condens. Matter **19**, 365216 (2007).

[80] Vikas, Chemical Physics Letters **458** 214 (2008).

[81] K. Higuchi and M. Higuchi, Phys. Rev. A **79**, 022113 (2009).

[82] M. Higuchi and K. Higuchi, Phys. Rev. A **81**, 042505 (2010).

[83] P. Blaha, K. Schwarz, G. K. H. Madsen, D. Kvasnicka and J. Luitz, in *WIEN2k*, ed. K. Schwarz, (Technische Universitat Wien, Vienna, 2001).

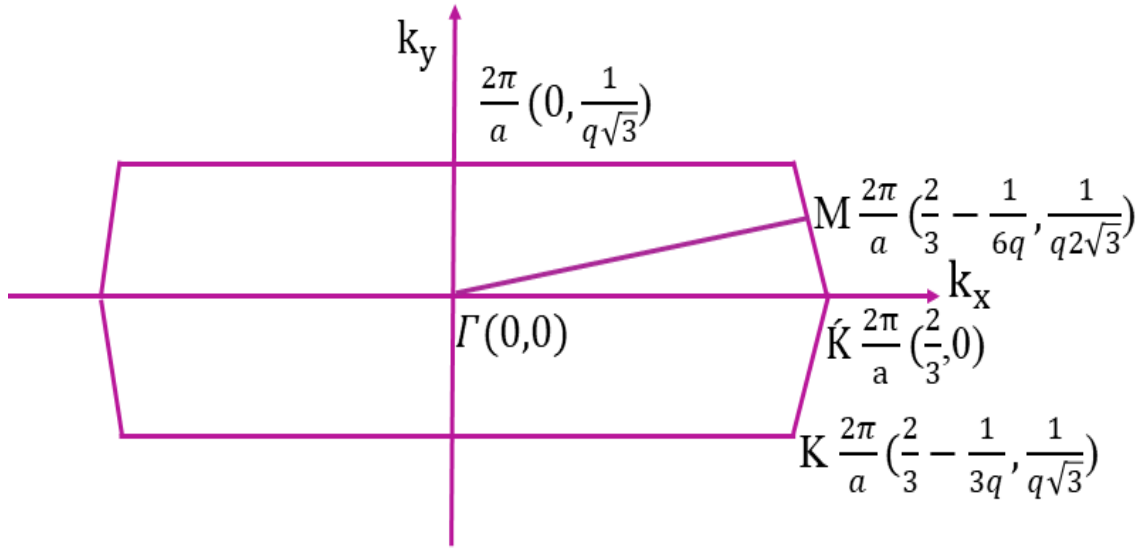

Fig. 1: Magnetic first Brillouin zone (MBZ) of graphene immersed in a magnetic field. The magnitude of the magnetic field is proportional to the rational number *p/q* (Eq. (1)).

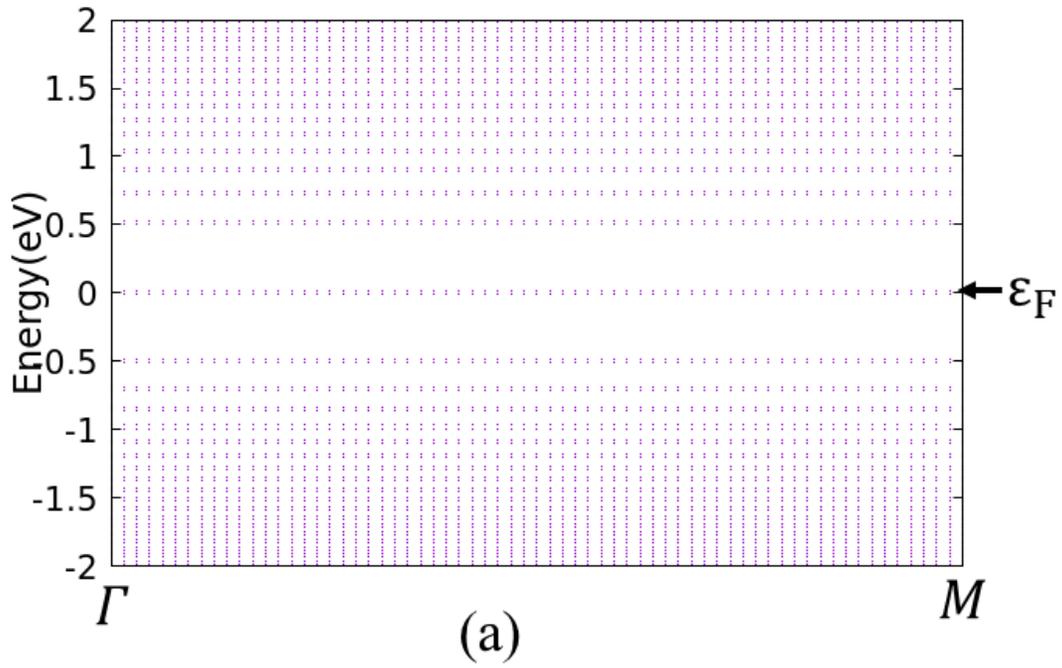

(a)

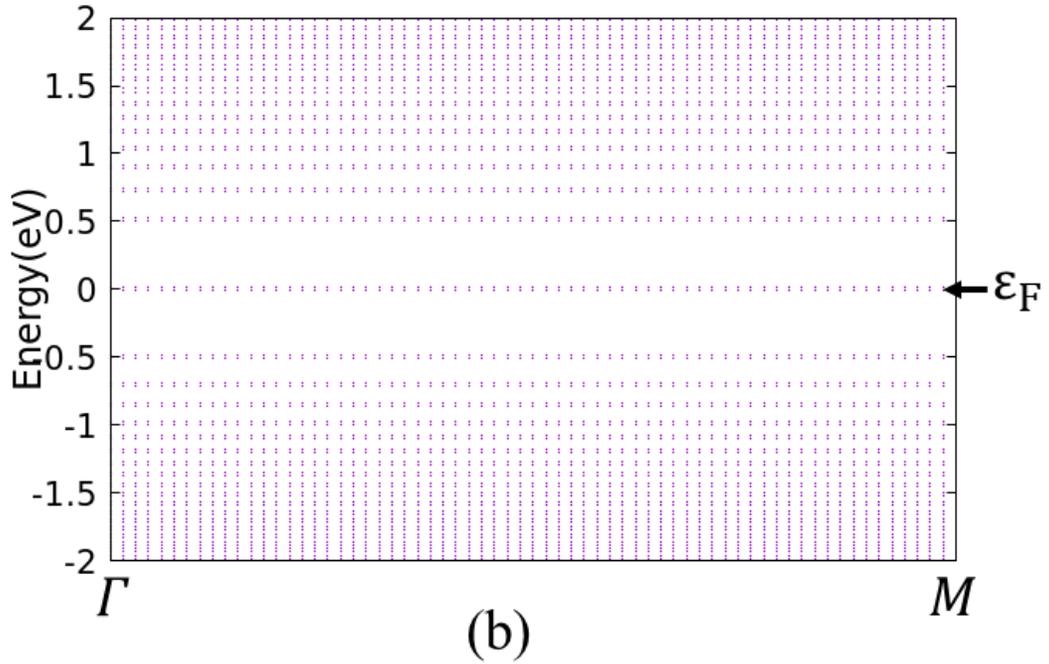

(b)

Fig. 2: Magnetic energy bands for (a) $B=200.5$ (T) and (b) $B'=200.9$ (T). The horizonal axis denotes the wave vector lying in the MBZ shown in Fig. 1. The arrow indicates the position of $\varepsilon_F$ for intrinsic graphene.

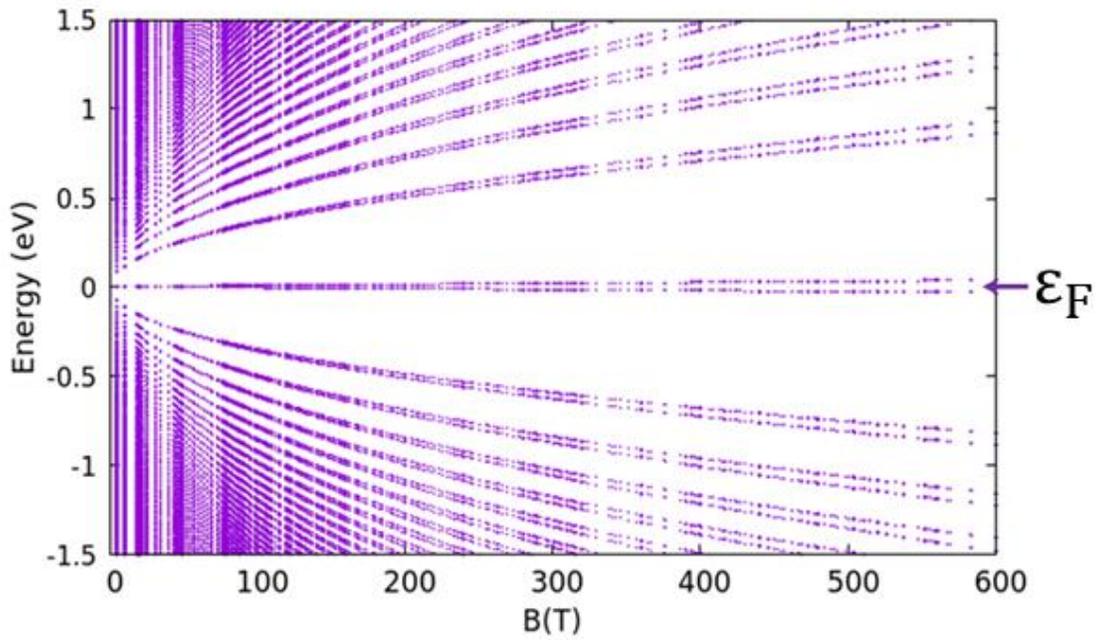

Fig. 3: Magnetic-field dependence of energy levels (Hofstadter butterfly diagram) calculated by the nonperturbative MFRTB method

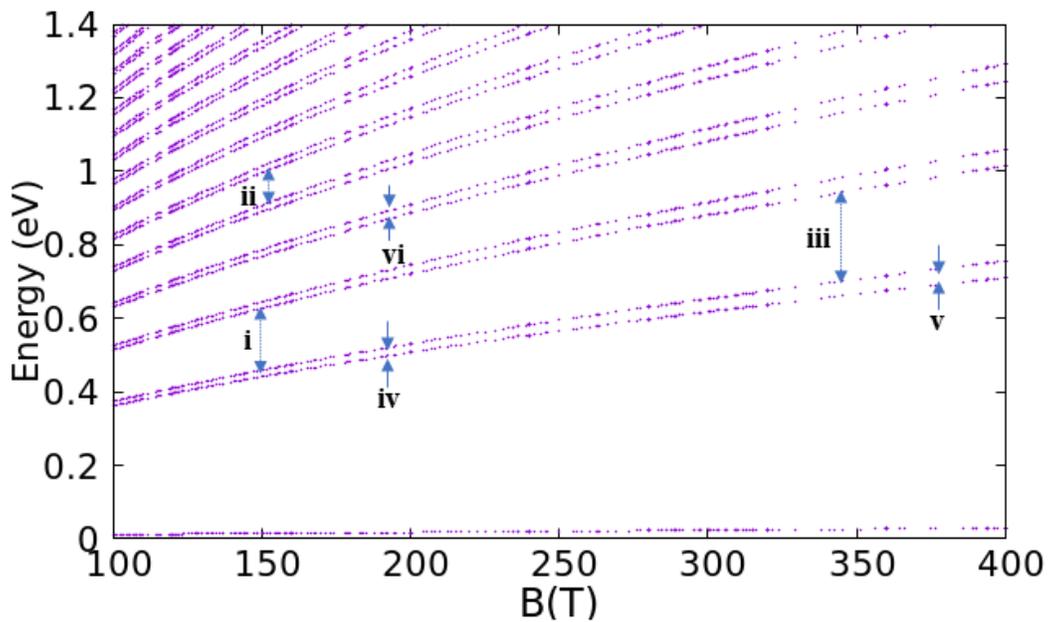

Fig. 4: Magnified view of Fig. 3

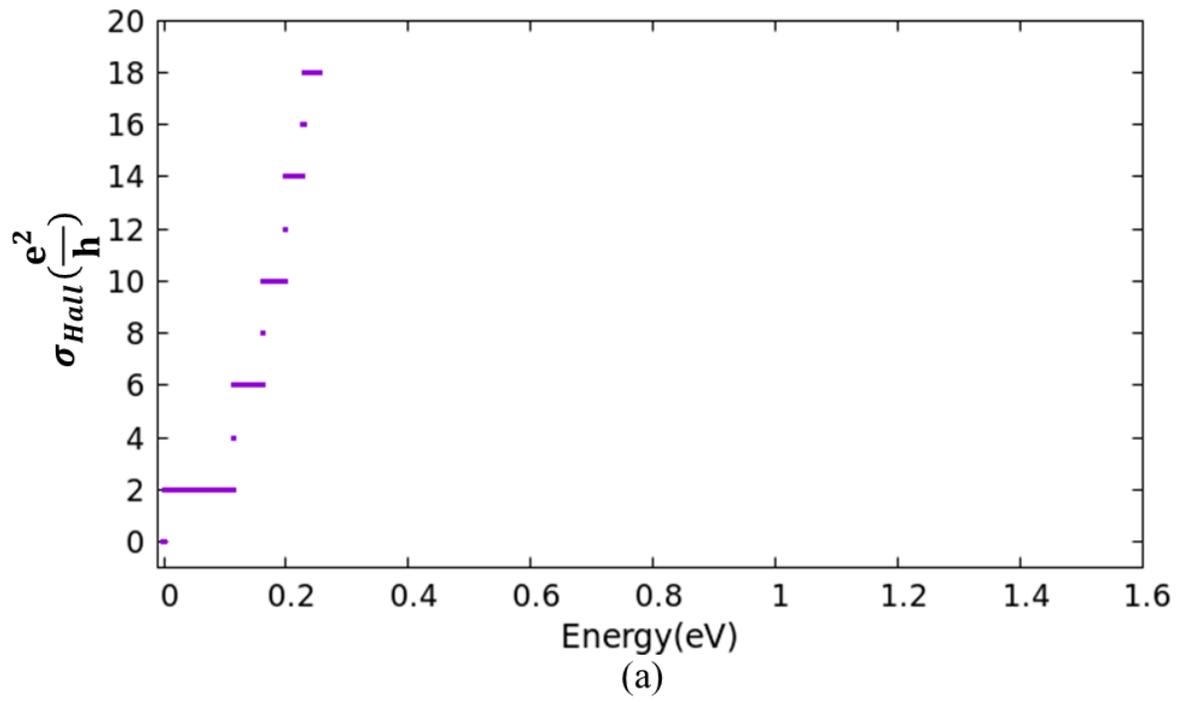

(a)

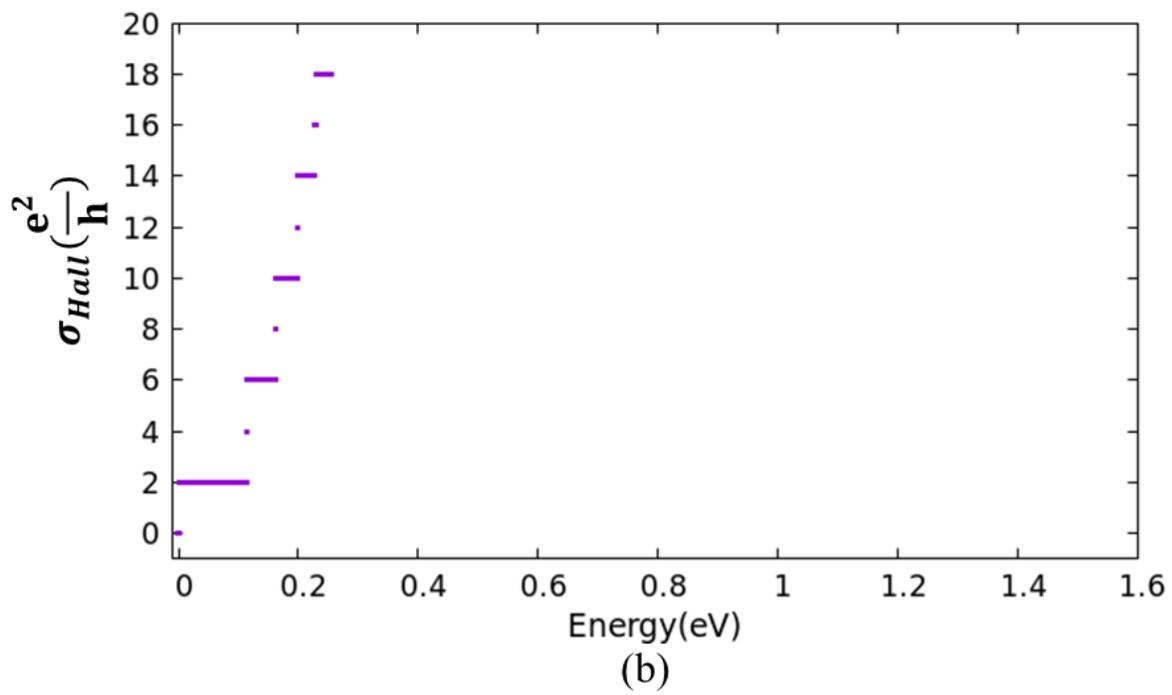

(b)

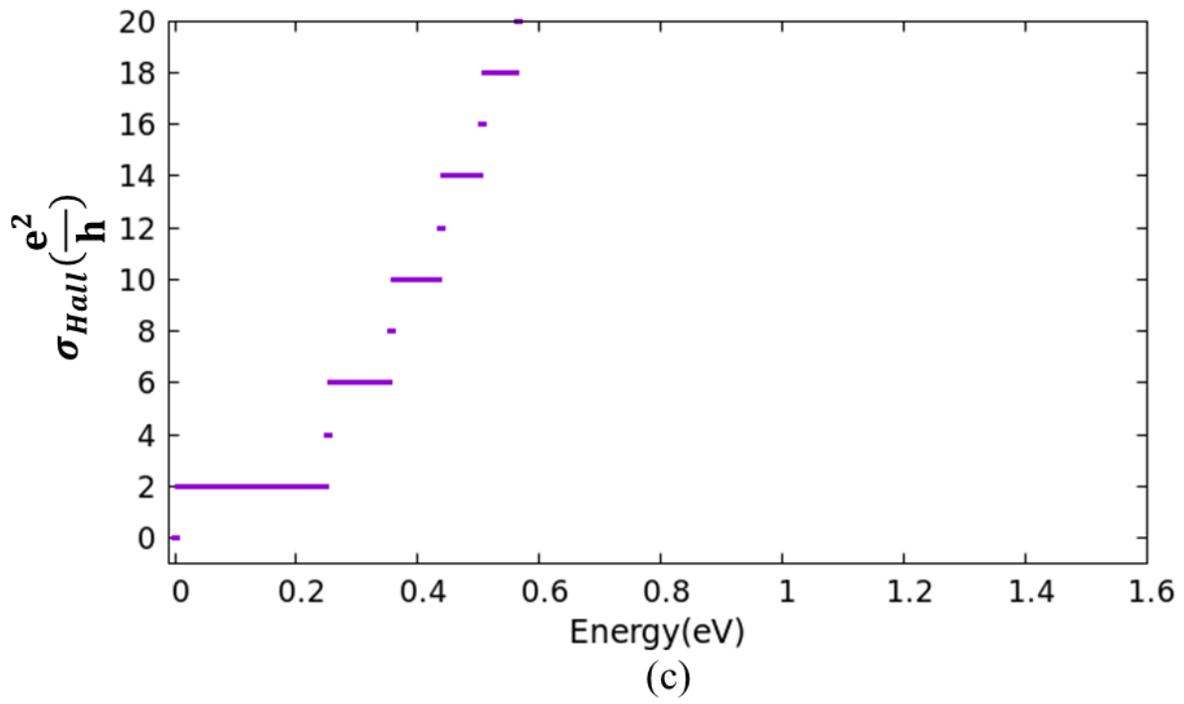

(c)

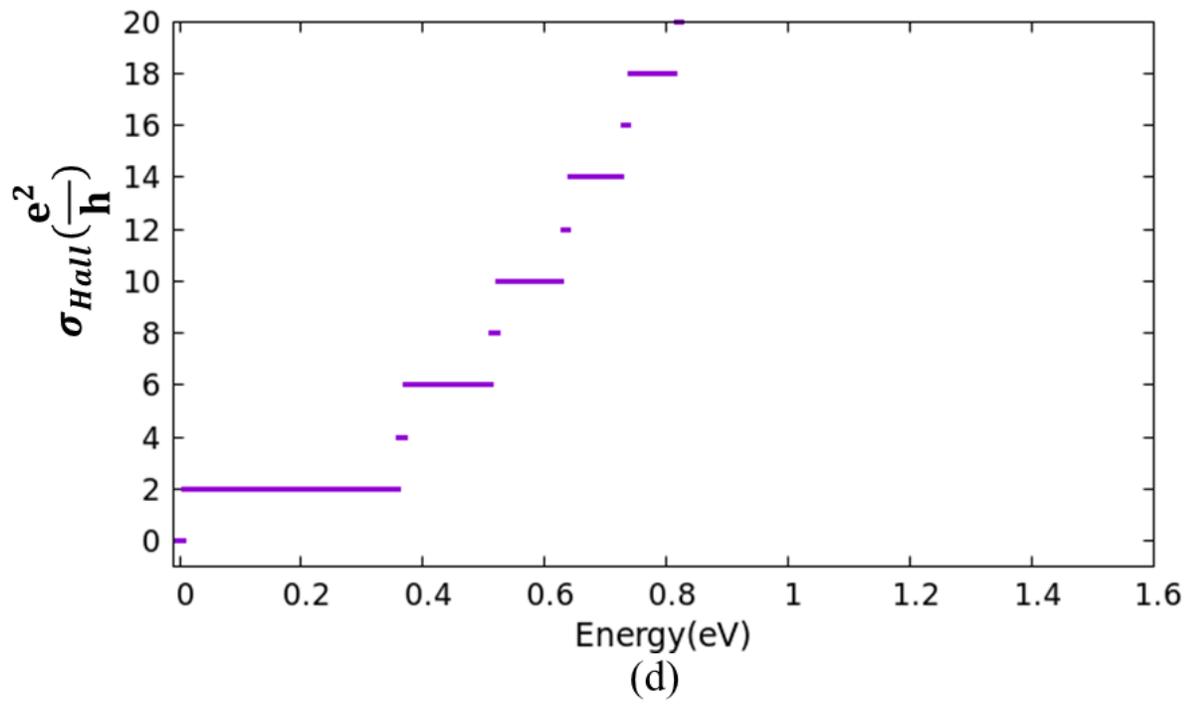

(d)

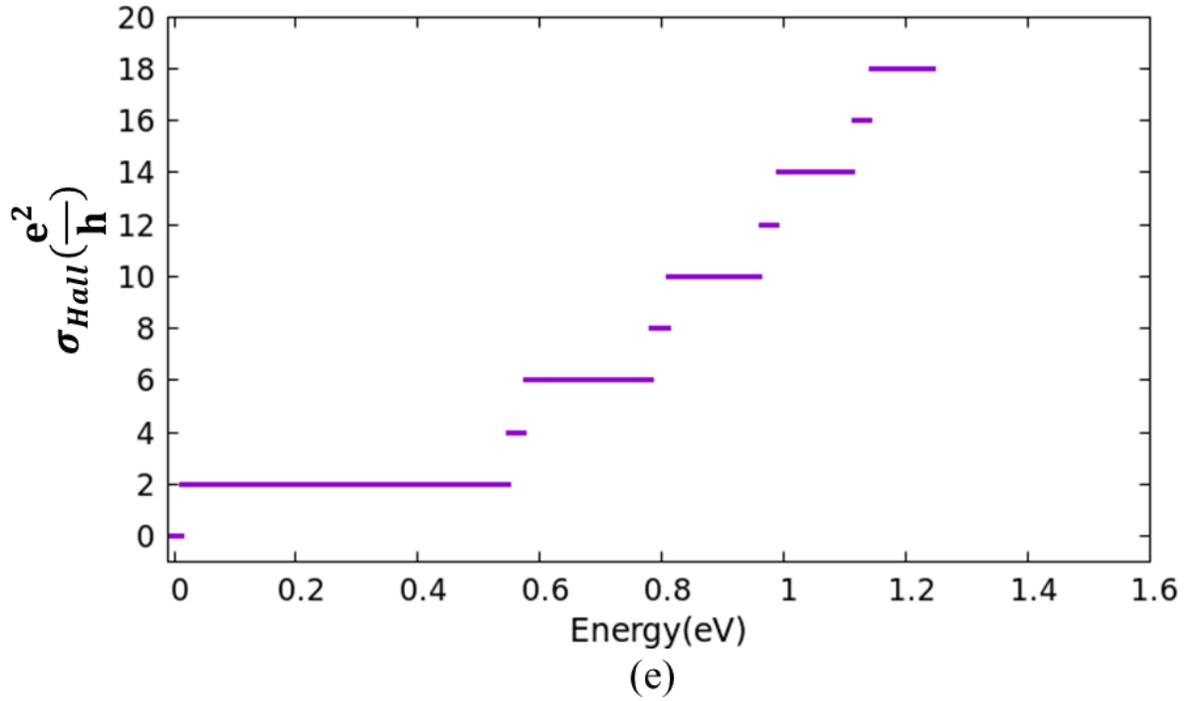

(e)

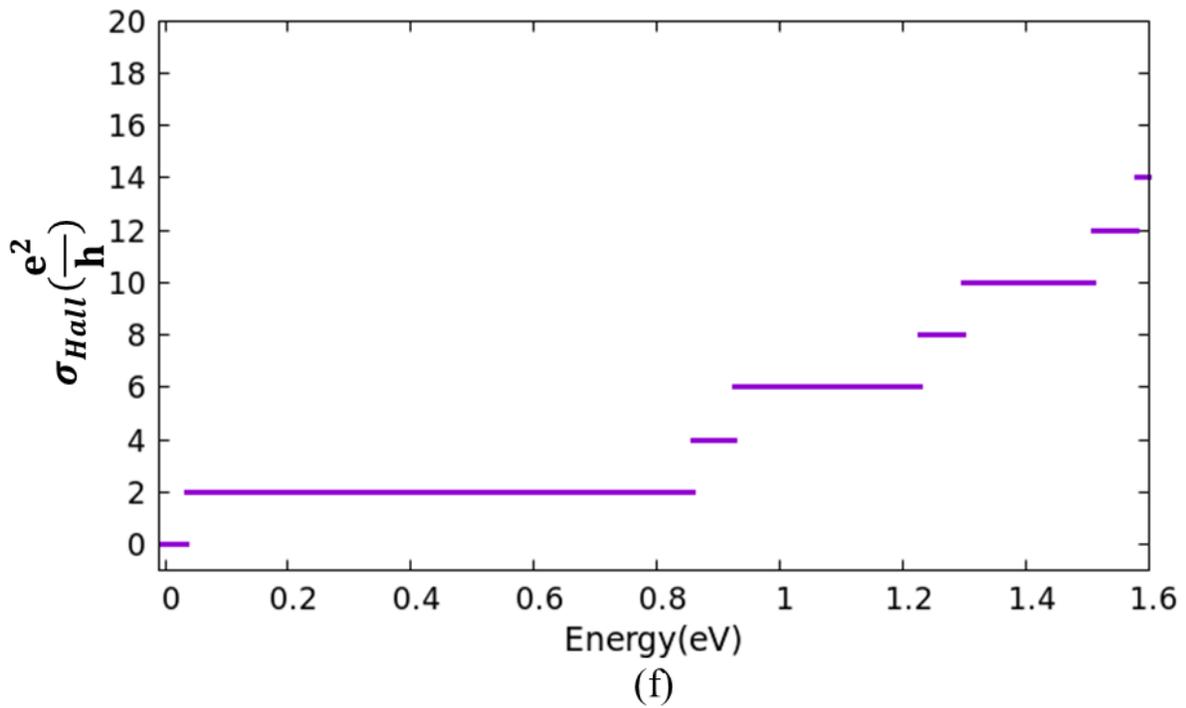

(f)

Fig. 5: Fermi energy dependence of the normalized Hall conductance ($\sigma_{Hall}(e^2/h)$) (a) for 10.00 (T), (b) for 20.02 (T), (c) for 48.50 (T) (d) for 101.50 (T), (e) for 239.50 (T) and (f) for 600.50 (T).

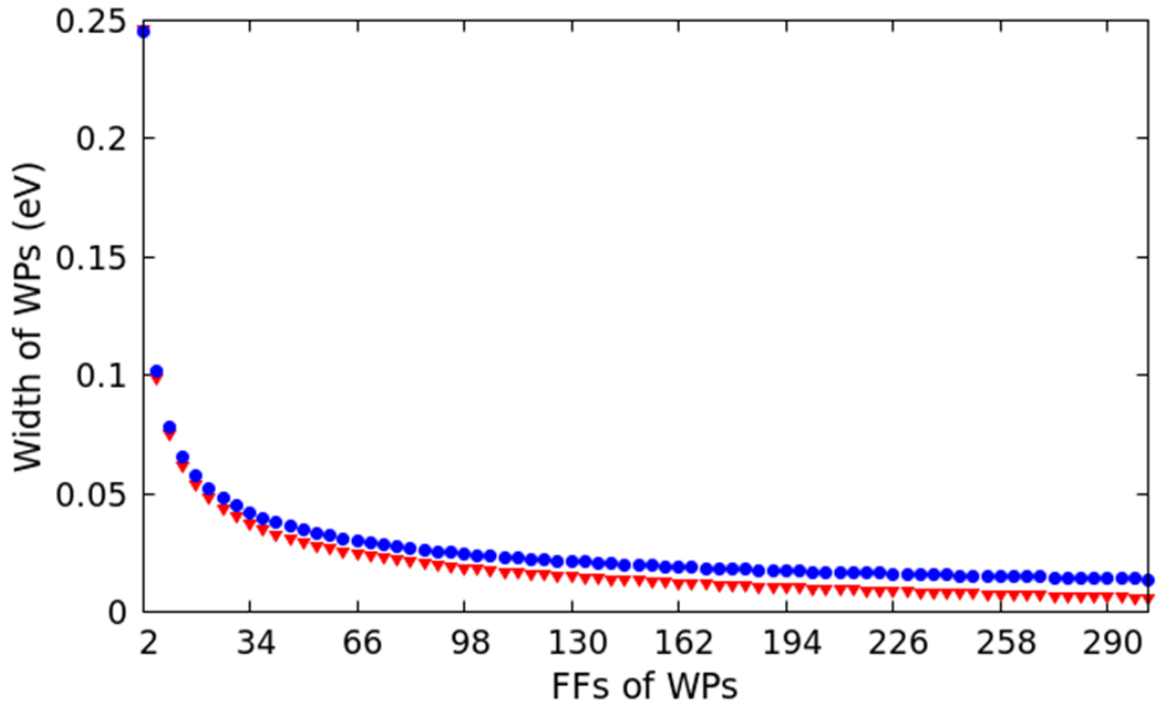

Fig. 6: Energy dependence of the width of WPs for 48.50 T, where solid circles (blue) represent the calculation from Eq. (5) and triangle (red) represents the calculation from the nonperturbative MFRTB.

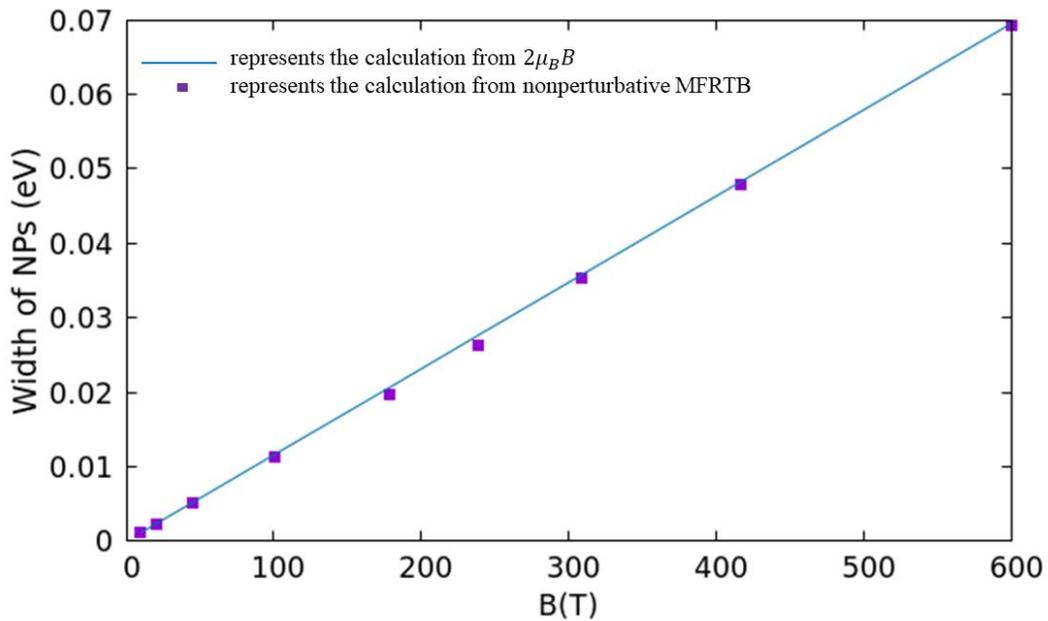

Fig. 7: Magnetic-field dependence of the width of the NP of FF=4. The width of the NP that is expected from the spin-Zeeman effect is also indicated by a solid line. The expected value is given by $2\mu_B B$, where $\mu_B$ denotes the Bohr magneton.

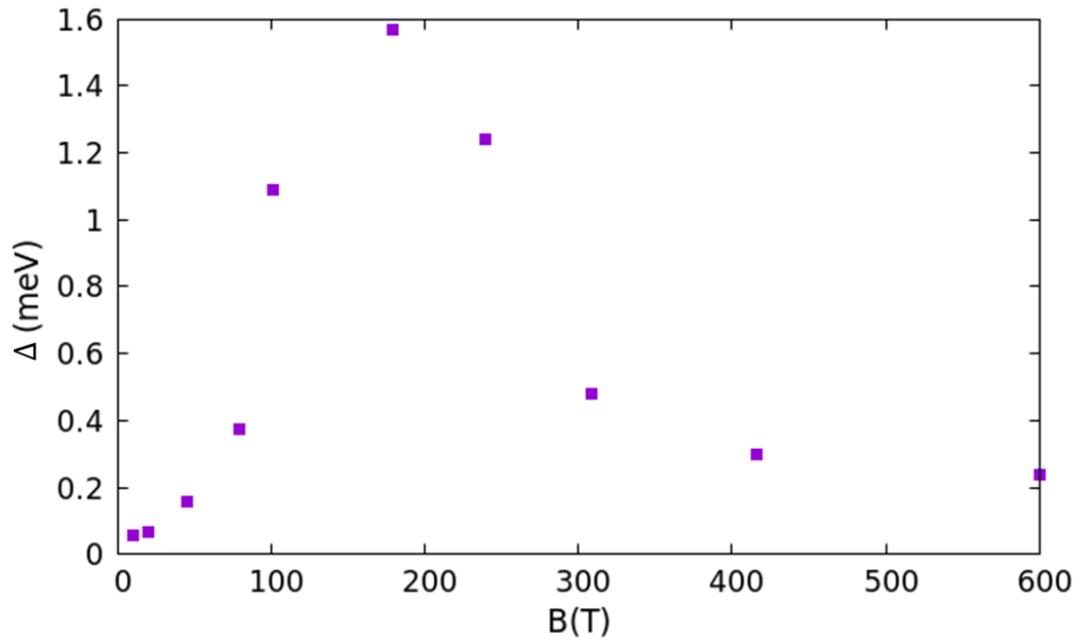

Fig. 8: Magnetic-field dependence of the difference ($\Delta$) between the width of the NP of FF=4 and the width expected from the spin-Zeeman effect.